\documentclass[12pt,a4]{article} %

\usepackage{amsthm}
\usepackage{amssymb}
\usepackage{graphicx,amsmath} 
\usepackage{mathtools} 
\usepackage{tikz}
\usepackage{natbib}
\usepackage{graphics}
\usepackage{subcaption}
\usepackage[justification=centering]{caption}
\usepackage{setspace}
\usepackage{tikz}
\usepackage[left=3cm,top=2.5cm,right=3cm,bottom=2.5cm]{geometry}
\usepackage{titlesec}
\usepackage{authblk}
\usepackage{multirow}
\usepackage{thmtools}
\usepackage[colorlinks=true,linkcolor=blue, urlcolor=blue, citecolor=blue, breaklinks]{hyperref}
\usepackage{thmtools}
\usepackage{pifont}

\titleformat{\section}{\filcenter \scshape}{\thesection.}{.5em}{}
\titleformat{\subsection}{\filcenter \small}{\thesubsection}{.5em}{}

\newtheorem{theorem}{Theorem}

\theoremstyle{definition}

\newtheorem{definition}{Definition}

\newcommand{\CC}{\mathcal{C}}
\newcommand{\BB}{\mathcal{B}}
\newcommand{\ccc}{\mathfrak{c}}
\newcommand{\bbb}{\mathfrak{b}}
\newcommand{\RR}{\mathbb{R}}
\newcommand{\PP}{\mathcal{P}}
\newcommand{\OO}{\mathcal{O}}
\newcommand{\ZZ}{\mathcal{Z}}
\newcommand{\SSS}{\mathcal{S}}
\newcommand{\UU}{\mathcal{U}}
\newcommand{\FF}{\mathcal{F}}
\newcommand{\TT}{\overline{T}}
\newcommand{\es}{\text{ES}} 
\newcommand{\IR}{\mathcal{I}}
\newcommand{\spaces}{\mathbb{\, \,}}
\definecolor{darkg}{rgb}{0.0, 0.5, 0.0}
\newcommand\floor[1]{\lfloor#1\rfloor}
\newcommand\ceil[1]{\lceil#1\rceil}

\begin{document}

\title{Equality of Opportunity and Integration in Social Networks\thanks{E-mail address: \href{mailto:ortega@zew.de}{ortega@zew.de}.}}

\author{Josu\'e Ortega}

\affil{Center for European Economic Research (ZEW), Mannheim, Germany.}

\date{\today}
\maketitle

\begin{abstract}
We propose the notion of $k$-integration as a measure of equality of opportunity in social networks. A social network is $k$-integrated if there is a path of length at most $k$ between any two individuals, thus guaranteeing that everybody has the same network opportunities to find a job, a romantic partner, or valuable information. We compute the minimum number of bridges (i.e. edges between nodes belonging to different components) or central nodes (those which are endpoints to a bridge) required to ensure $k$-integration. The answer depends only linearly on the size of each component for $k=2$, and does not depend on the size of each component for $k \geq 3$. Our findings provide a simple and intuitive way to compare the equality of opportunity of real-life social networks.
\end{abstract}
\vspace*{1cm}

 {\small {\sc KEYWORDS}: social integration, social networks, equality of opportunity.}

\vfil{}
\thispagestyle{empty}
\pagebreak

\section{Relevance}

Our social networks are highly clustered, meaning that it is common to observe communities that are very densely connected within themselves but only poorly so with the rest of the social network \citep{watts1998,traud2012,mishra2007}. The people in each community usually share particular characteristics, such as race, ethnicity or educational background. For example, the average American public school student has less than one school friend of another race \citep{fryer2007}. Among White Americans, 91\% of people comprising their social networks are also White \citep{values2}. Among British students who live in highly multicultural residences, less than 5\% nominates a South-American, West Indian, Oriental or Middle Easterner among their three best friends \citep{furnham1985friendship}. Social networks are also highly clustered by religion, political views, and education, among other characteristics \citep{louch2000}.

The fact that we are segregated into these poorly interconnected societies has significant consequences. Many important decisions such as who we marry, who we work for, and what information we obtain crucially depend on who we know or who our friends and our acquaintances know. We are incredibly likely to marry someone within our extended social circles (at least until the emergence of online dating, see \citealp{rosenfeld2012} and \citealp{ortega2017}). It is also well-documented that over 50\% of jobs are obtained through social contacts \citep{rees1966, corcoran1980, granovetter1973, granovetter1995}, and that the connections we possess may explain the observed drop-out rates of the labor force \citep{topa2001, calvo2004}. Furthermore, our social circle determines the information to which we are exposed to online \citep{bashky2012}, our sexual partners \citep{ergun} and how scientists collaborate in their research projects \citep{barabasi2002,farkas2002}. Therefore, the lack of connections between communities can forbid that some agents interact with others, making them lose potential job opportunities, valuable information, or potential romantic partners.

In this paper, we study the minimum number of inter-community connections required so that every agent in the social network has access to the entire social network, thus guaranteeing that every agent has the same networking opportunities. To do so, we assume that each agent can only interact with another agent who is connected to her within $k$ steps, i.e. connected by a path of length at most $k$. For example, $k=1$ means that agents can only interact (work, marry, communicate, etcetera) with people whom they directly know. Similarly, $k=2$ means that agents can interact with people who they either know directly, or with whom they have a friend in common, and so on.

Given $k$, the question we ask is how many connections between communities (which we call {\it bridges}) are required to guarantee that any two people in the social network are $k$-connected, i.e. that the underlying graph has diameter $k$? And how many agents must know someone outside their community, i.e. how many agents must be {\it central}? In this paper, we provide answers to both questions for all values of $k$ as a function of the number and the size of the communities conforming a social network. We obtain tight values, which guarantee that if we remove one bridge or one central agent, there must be at least two nodes that are not connected by a path of length $k$. In this case, we say that the social network fails to be $k$-integrated. Because we study lower bounds on the number of bridges and nodes, we impose without loss of generality that every community is completely connected within, i.e. the corresponding graph of each community is complete.

The notion of $k$-integration for social networks that we propose, which requires that any two nodes are connected by a path of at most length $k$, is inspired by the well-known literature of equality of opportunity in philosophy and economics \citep{roemer2009,arneson2015}. If a social network is $k$-integrated, every agent would have the same access to job opportunities, information, romantic partners, and so on, irrespective of their location within the network and the community to which the agent belongs. Our results allow us to identify when a social network fails our proposed criterion of opportunity fairness in networks, which is parametrized by $k$, simply by either counting the number of connections between communities or the number of agents who have such connections. It also allows us to compare social networks on the basis of $k$-integration.

\begin{figure}[!ht]
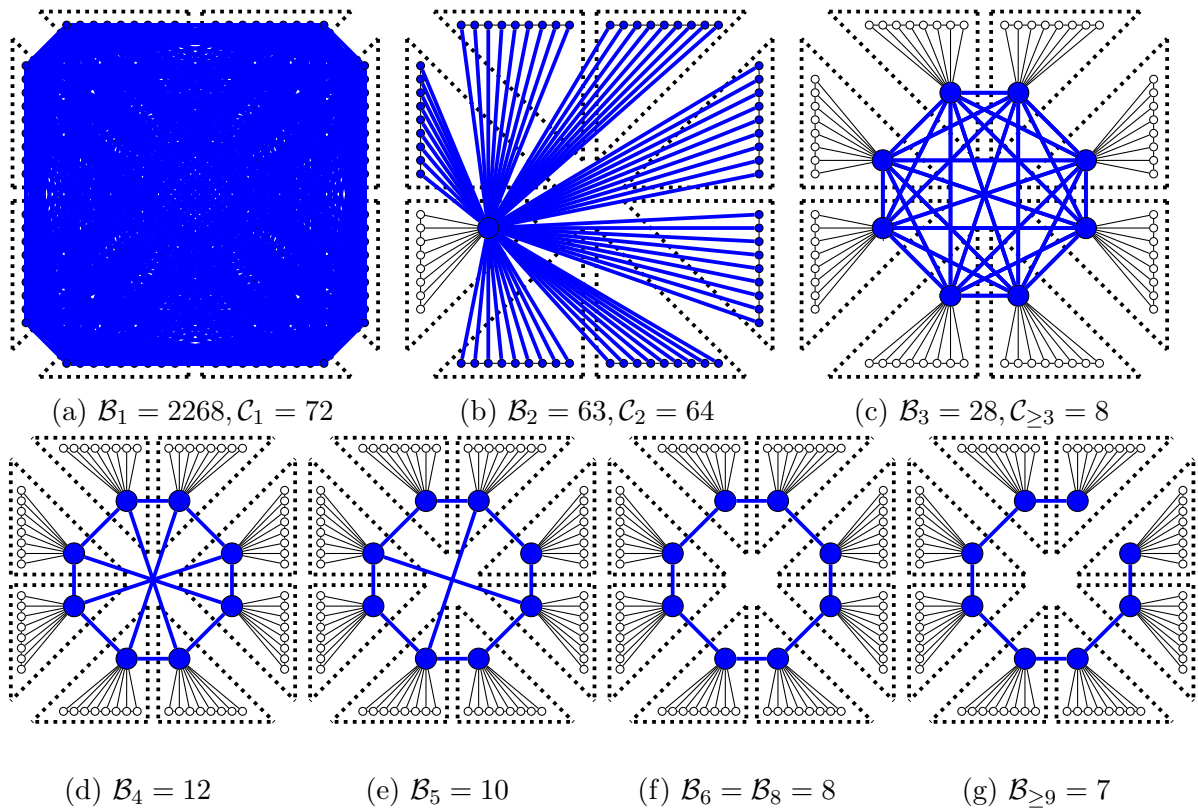

    \centering
    \begin{subfigure}[b]{0.31\textwidth}
\centering


        \caption{$\BB_1=2268, \CC_1=72$}
					\label{fig:1}
    \end{subfigure}
    ~ 
    \begin{subfigure}[b]{0.31\textwidth}
        \centering
        %

        \caption{$\BB_2=63, \CC_2=64$}
				\label{fig:2}
    \end{subfigure}
    ~ 
    \begin{subfigure}[b]{0.31\textwidth}
				\centering
				%

        \caption{$\BB_3=28, \CC_{\geq 3}=8$}
					\label{fig:3}
    \end{subfigure}

		\begin{subfigure}[b]{0.22\textwidth}
        \begin{center}    
        %
\end{center}
        \caption{$\BB_4=12$}
        					\label{fig:4}
    \end{subfigure}
    \quad 
    \begin{subfigure}[b]{0.22\textwidth}
      \begin{center}
      	%
      \end{center}

        \caption{$\BB_5=10$}
        					\label{fig:5}
    \end{subfigure}
    \quad 
    \begin{subfigure}[b]{0.22\textwidth}
				\begin{center}
					%
				\end{center}
				
        \caption{$\BB_6=\BB_8=8$}
        					\label{fig:6}
    \end{subfigure}
		\quad 
    \begin{subfigure}[b]{0.22\textwidth}
				\begin{center}
					%
				\end{center}

        \caption{$\BB_{\geq 9}=7$}
					\label{fig:7}
    \end{subfigure}

    \caption{Minimal $k$-integrated social networks with 8 communities of 9 agents each. {\footnotesize Bridges and central nodes appear in blue. Communities are indicated with dashed triangles. The size and location of nodes are irrelevant and for exposition only. Each community has a complete graph: some of these intra-community edges do not appear to improve readability.}}
            \label{fig:uno}
\end{figure}

We find that the number of bridges ($\BB_k$) and central agents ($\CC_k$) needed for $k$-integration quickly diminishes as $k$ becomes larger, and does not depend on the size of each community for $k \geq 3$. To give an example, consider a social network with 8 communities, each containing 1,000 agents. For $1-$integration, we require 28 million bridges and 8,000 central agents, whereas for $2-$integration we only require 7,000 bridges and 7,001 central agents. Furthermore, for $3-$integration we only require 28 bridges and 8 central agents. The intuition behind our result is that as $k$ grows, the externalities generated by each bridge and central node increase. It is nevertheless surprising that the externalities generated by each bridge are so large that, to cover such a large social network with $k=3$, we only need 0.00010\% of the total possible bridges and 0.1\% of the total nodes to be central. Figure \ref{fig:uno} provides an illustration of a social network with 8 communities with 9 agents each.

Our result is related to the small world phenomenon in \citet{watts1998}. They note that regular graphs (those in which all nodes have the same number of neighbors) are highly clustered but exhibit long characteristic path lengths. On the contrary, random graphs in which edges are created independently at random, have short characteristic path lengths but are not clustered. However, if we start from a regular graph and we randomly rewire each edge with some probability, we quickly obtain a graph that is both highly clustered and exhibit short characteristic path lengths, two properties of many real-life networks.\footnote{\cite{barthelemy1999small} show that the appearance of small-world behavior is not a phase-transition but a crossover phenomenon.} Our result is similar to theirs in spirit, but instead of looking at characteristic or expected path lengths, we strictly require that each path length is at most $k$. Although a random graph may exhibit a short characteristic path length, it may still have a few edges that are very poorly connected to the rest of the social network. Thus, our integration measure ensures that all agents are well-connected to the network, unlike expected or characteristic path length. We quantify the specific number of bridges and central nodes needed to ensure that all paths are of length at most $k$.

\cite{ortega2017} also obtain a similar result using random graphs, which they use to analyze the impact of online dating in the number of interracial marriages. They show that for $k \geq 2$, a few bridges can quickly make a social network 2-integrated, leading to a fast increase in the number of agents who decide to marry interracially. Their model's prediction is consistent with observed demographic trends. In this paper, we focus instead on deterministic graphs and compute the minimal number of bridges and central nodes required to guarantee $k$-integration for all values of $k$. This paper contributes to explaining the fast increase of interracial marriages in their model, by showing that just a few bridges can ensure that the society is $k$-integrated for $k\geq 2$. Furthermore, it provides an easy way to check whether a social network may experience such an integration change.\footnote{In general, making a graph more connected can reduce some desirable properties of the graph \citep{ortega2, ortega}.}

There is a vast literature on efficiently computing the exact value of $k$ (i.e the diameter) in arbitrary social networks, which is a substantially more general problem; see \cite{aingworth} and references therein.

\section{Definitions and Model}

{\it Basic Definitions.} An undirected {\bf graph} or {\bf network}\footnote{We use both terms indistinctly.} is an ordered pair $F=(N,E)$ consisting of a set of {\bf nodes} $N=\{a,b,c, \ldots\}$ and a set of {\bf edges} $E=\{ab,ac,\ldots\}$. The set of edges consists of unordered pairs of nodes. Two nodes $a, b$ are {\bf adjacent} if $ab \in E$, and we say that $a$ and $b$ are the {\bf endpoints} of edge $ab$. A graph in which every two different nodes are adjacent is called {\bf complete}. A graph $F'=(N',E')$ is a {\bf supergraph} of $F$ if $N \subseteq N'$ and $E \subseteq E'$. Two graphs are {\bf disjoint} if they share no nodes. A {\bf path} from node $i$ to $t$ is a vector of edges $(ij,jk,\ldots,qs,st)$ such that every consecutive edges share an endpoint. The {\bf path length} is the number of edges in a path. The {\bf distance} between two nodes is the length of the shortest path between them. Two graphs are {\bf connected} if there exists a path between any of their nodes, and disconnected otherwise. 

{\it Model.} We consider $r$ disjoint complete graphs $K_1=(N_1,E_1), K_2=(N_2,E_2),\ldots, K_r=(N_r,E_r)$, all with $n$ nodes. Each complete graph is called a {\bf local graph}, and its edges are called {\bf local edges}. A {\bf global graph} $G=(N_G,E_G)$ is a supergraph of the union of all local graphs, so that $\bigcup\limits^{r}_{i=1} N_i \subseteq N_G$ and $\bigcup\limits_{i=1}^r E_i \subseteq E_G$. The notation $K_i \cup K_j$ denotes the supergraph $(N_i \cup N_j, E_i \cup E_j)$. We assume $n\geq r$. A supergraph is equivalent to the islands network structure proposed by \cite{golub2012} which, in their words, {\it ``is surprisingly accurate as a description of some friendship patterns''}.

A {\bf bridge} is an edge of a global graph whose endpoints belong to different local graphs.\footnote{The standard definition of a bridge is an edge whose removal would disconnect the graph \citep{diestel2017}. Nevertheless, we redefine this term to make exposition clearer.} A node is {\bf central} if its an endpoint of a bridge. $\BB$ and $\CC$ denote the number of bridges and central nodes in a global graph, respectively. Figure \ref{fig:example} presents an example of a global graph with two bridges and four central nodes.

\begin{figure}[!htbp]
	\centering
	\begin{center}
		\begin{tikzpicture}
		\node[circle,draw=black, inner sep=0pt,minimum size=16pt] (01) at (-3,-1) {};
		\node[circle,draw=black, inner sep=0pt,fill=blue,   minimum size=16pt] (02) at (-2,-1) {};
		\node[circle,draw=black, inner sep=0pt,minimum size=16pt] (03) at (-2,-2) {};
		\node[circle,draw=black, inner sep=0pt,minimum size=16pt] (04) at (-3,-2) {};
		
		\node[circle,draw=black, inner sep=0pt,minimum size=16pt] (11) at (-1,2) {};
		\node[circle,draw=black, inner sep=0pt,fill=blue,  minimum size=16pt] (12) at (-1,1) {};
		\node[circle,draw=black, inner sep=0pt,fill=blue,  minimum size=16pt] (13) at (0,1) {};
		\node[circle,draw=black, inner sep=0pt,minimum size=16pt] (14) at (0,2) {};
		
		\node[circle,draw=black, inner sep=0pt,minimum size=16pt] (21) at (1,-2) {};
		\node[circle,draw=black, inner sep=0pt,fill=blue,  minimum size=16pt] (22) at (1,-1) {};
		\node[circle,draw=black, inner sep=0pt,minimum size=16pt] (23) at (2,-1) {};
		\node[circle,draw=black, inner sep=0pt,minimum size=16pt] (24) at (2,-2) {};
		
		\foreach \z in {0,1,2}
		\foreach \x in {1,...,4}
		\foreach \y in {2,...,4}  
		\draw (\z\x)--(\z\y);
		\draw[line width=0.5mm, blue ] (12)--(02);
		\draw[line width=0.5mm, blue ] (13)--(22);
		
		\draw[dotted,line width=0.5mm] (-.5,1.5) circle (1);
		\draw[dotted,line width=0.5mm] (-2.5,-1.5) circle (1);
		\draw[dotted,line width=0.5mm] (1.5,-1.5) circle (1);
		\end{tikzpicture}
		\caption{A global graph with $n=4$, $r=3$, $\BB=2$ and $\CC=4$.}
		\label{fig:example}
	\end{center}
\end{figure}

We propose the following notion of integration for global graphs.

\begin{definition}[Integration in Graphs]
	A global graph is $k$-integrated if there exists a path of length at most $k$ between any two nodes. A global graph that is not $k$-integrated is $k$-segregated.
\end{definition}

This paper studies the minimum number of bridges and central agents that a global graph or network requires to be $k$-integrated. These are presented in the following Theorem.

\section{Result}

\begin{theorem}
A global graph is $k$-segregated if $\BB<\BB_k$ or $\CC<\CC_k$. The values for $\BB_k$ and $\CC_k$ are given in the table below and are tight, i.e. for any values of $r$ and $n$ there exist a $k$-integrated global graph with $\BB=\BB_k$ and $\CC=\CC_k$.

\begin{table}[!htbp]
\centering
\begin{tabular}{lll}
\hline
$k$ & $\BB_k$ & $\CC_k$ \\
\hline
1 & $n^2r(r-1)/2$ & $rn$ \\
2 & $(r-1)n$ & $(r-1)n+1$ \\
3 & $r(r-1)/2$ & $r$ \\
\vdots & $f(r)$ & $r$ \\
$r+1$ & $r-1$ & $r$ \\
\hline
\end{tabular}
\end{table}

where $f(r)$ is a function of $r$ that does not depend on $n$.
\end{theorem}

We separate the proof into three parts. First, we show that a global graph is $k$-segregated if $\BB<\BB_k$. Second, we show that a global graph is $k$-segregated if $\CC<\CC_k$. Third, we show that one can always construct a $k$-integrated local graph with exactly $\BB_k$ bridges and $\CC_k$ central nodes, for any values of $r$ and $n$.

\begin{proof}[Proof of Minimality of $\BB_k$]
For $k=1$, order all local graphs randomly. Each node in the first local graph needs to be adjacent to $n$ nodes in each of the $(r-1)$ local graphs. Each node in the second local graph needs to be connected to $n$ nodes in each of the remaining $(r-2)$ local graphs, and so on. A simple computation yields
	\begin{equation}
	n \sum_{i=1}^{r-1} i\cdot n=n^2\frac{r(r-1)}{2}
	\end{equation}

For $k=2$, we start by showing the following Lemma: Let $A=(N_1,E_1)$ and $B=(N_2,E_2)$ be two disjoint graphs with $n_1$ and $n_2$ nodes. Then $\BB_k=\min(n_1,n_2)$. Otherwise, there are two nodes $x \in N_1$, $y \in N_2$ such that both are not central. Thus, it is impossible to make a path of length at most 2 connecting $x$ and $y$.

Now we show that the global graph must have $\BB_2=(r-1)n$ by constructing such a graph. Start with a global graph containing all $r$ local graphs $K_1, K_2, \ldots, K_r$ and no bridges. To ensure that every node in $K_1$ is 2-integrated to $K_2$ we need $n$ bridges, by the previous Lemma. Similarly, we need $n$ bridges to 2-integrate $K_1 \cup K_2$ and $K_3$ (remember that we started from a graph with 0 bridges), $n$ bridges to 2-integrate $K_1 \cup K_2 \cup K_3$ and $K_4$, ..., and $n$ bridges to 2-connect $\bigcup\limits_{i=1}^{r-1} K_{i}$ and $K_r$. This is the minimal number of bridges we need to construct to ensure that the graph is 2-connected. In fact, if the bridges are created in a particular way, we do not need any bridges connecting any two local graphs $K_i, K_j$ for $i \neq j \neq 1$. We show this when presenting an example showing that this bound is tight in part 3 of the proof.

For $k=3$, assume that two local graphs $K_1$ and $K_2$ are not directly connected by a bridge, i.e. $\nexists x \in K_1, y \in K_2$ such that $xy \in \BB$. Then, for that graph to be 3-integrated, all the $n$ nodes of $K_1$ must be directly connected to a local graph $K_3$ that is connected to $K_2$ by a bridge. Thus, for each two local graphs that are not directly connected, we need $n$ bridges. Now suppose that all but one local graph are directly connected by a bridge. The total number of bridges is $n+\frac{(r-1)(r-2)}{2}$, whereas if every two local graphs were directly connected by a bridge the total number of bridges would be $\frac{r(r-1)}{2}$. A quick computation makes evident (recall $n \geq r$) that
\begin{eqnarray}
\frac{r(r-1)}{2} < n+\frac{(r-1)(r-2)}{2}
\end{eqnarray}

Finally, for $k>3$, note that $\BB_k \geq \BB_{k+1}$. Therefore, $\BB_k \leq \frac{r(r-1)}{2}$ for $k>3$, which is itself a function of $r$ which does not depend on $n$. Finally, note that for $k=r+1$, we need at least a path between any two nodes. This means that all but two local graphs must be directly connected by a bridge to at least one other local graph. If $\BB_{r+1}<r-1$, then there are two local graphs that are not connected by any path, a contradiction. This concludes the proof of minimality of $\BB_k$.
\end{proof}

Now we show that a global graph is $k$-segregated if $\CC<\CC_k$.

\begin{proof}[Proof of Minimality of $\CC_k$]
For $k=1$ every agent needs to be central so the proof is trivial. 

For $k=2$, suppose by contradiction that there are only $(r-1)n$ central nodes in a 2-integrated global graph, i.e. there are $n$ non-central nodes. Then either all those non-central nodes belong to a whole local graph, which is then disconnected from the rest of the global graph (a contradiction), or there are two nodes from different local graphs that are not central. The shortest path between those has length 3, a contradiction.

For $k\geq 3$ the proof is also trivial: if there are less than $r$ central nodes then at least one local graph is disconnected from the rest of the global graph.
\end{proof}

Finally, we provide examples of global graphs that are $k$-integrated with exactly $\BB=\BB_k$ and $\CC=\CC_k$, showing that these values are tight. 

\begin{proof}[Proof of Sufficiency of $\BB_k$ and $\CC_k$]
For $k=1$, we trivially require that every node is central and that any two nodes belonging to different communities are directly connected by a bridge (see subfigure \ref{fig:1} in the Introduction).

For $k=2$, we can achieve the lower bounds $\BB_2=(r-1)n$ and $\CC_2=(r-1)n+1$ by selecting one local graph that has only one central node. This node connects to all other nodes of different communities, and by our assumption also to every node in his own community. This way we create a {\it``2-integrated star graph''} which is by construction 2-integrated. An example appears in Figure \ref{fig:example3} (and in subfigure \ref{fig:2} in the Introduction).

\begin{figure}[ht]
	\centering
	\begin{center}
		\begin{tikzpicture}
		\node[circle,draw=black, inner sep=0pt,minimum size=16pt] (01) at (-5,0) {};
		\node[circle,draw=black, inner sep=0pt,fill=blue,  minimum size=16pt] (02) at (-3,0) {};
		\node[circle,draw=black, inner sep=0pt,minimum size=16pt] (03) at (-5,-1) {};
		
		\node[circle,draw=black, inner sep=0pt,fill=blue,  minimum size=16pt] (11) at (-1,1) {};
		\node[circle,draw=black, inner sep=0pt,fill=blue,  minimum size=16pt] (12) at (0,1) {};
		\node[circle,draw=black, inner sep=0pt,fill=blue,  minimum size=16pt] (13) at (1,1) {};
		
		\node[circle,draw=black, inner sep=0pt,fill=blue,  minimum size=16pt] (21) at (2,0) {};
		\node[circle,draw=black, inner sep=0pt,fill=blue,  minimum size=16pt] (22) at (2,-1) {};
		\node[circle,draw=black, inner sep=0pt,fill=blue,  minimum size=16pt] (23) at (2,-2) {};
			
		\foreach \z in {0}
		\foreach \x in {1,...,3}
		\foreach \y in {2,...,3}  
		\draw (\z\x)--(\z\y);

		\foreach \i in {2}
		\foreach \x in {1,2}
		\foreach \y in {1,...,3}  
		\draw[line width=0.5mm, blue ] (0\i)--(\x\y);
		
		\draw (13)--(12);
		\draw (11)--(12);
		\draw (23)--(22);
		\draw (21)--(22);
		
		\draw    (13) to[out=150,in=30] (11);
		\draw    (23) to[out=60,in=-60] (21);

		\draw[dotted,line width=0.5mm] (-4,-.5) ellipse (60pt and 35pt);
		\draw[dotted,line width=0.5mm] (2,-1) ellipse (15pt and 45pt);
		\draw[dotted,line width=0.5mm] (0,1) ellipse (45pt and 15pt);
		\end{tikzpicture}
				\caption{A 2-integrated star graph with $\BB_2=(r-1)n$ and $\CC_2=(r-1)n+1$.}
		\label{fig:example3}
	\end{center}
\end{figure}

For $k=3$, we can achieve the lower bounds $\BB_3=r(r-1)/2$ and $\CC_3=r$ by selecting one node from each local graph, which will be the only central node belonging to his community. This node is connected to all other central nodes of each other local graph. Therefore, we create a complete graph among each of the $r$ central nodes. We call the corresponding global graph an {\it ``extended star graph''}. An example appears in Figure \ref{fig:example4}.

\begin{figure}[ht]
\centering
\begin{center}
	\begin{tikzpicture}
	\node[circle,draw=black, fill=blue,   inner sep=0pt,minimum size=16pt] (c1) at (1,1) {};
	\node[circle,draw=black, inner sep=0pt,minimum size=16pt] (c2) at (2,1) {};
	\node[circle,draw=black, inner sep=0pt,minimum size=16pt] (c3) at (1.5,1.5) {};
	\node[circle,draw=black, inner sep=0pt,minimum size=16pt] (c4) at (1,2) {};
		
	\node[circle,draw=black, fill=blue,   inner sep=0pt,minimum size=16pt] (c11) at (-1,1) {};
	\node[circle,draw=black, inner sep=0pt,minimum size=16pt] (c12) at (-2,1) {};
	\node[circle,draw=black, inner sep=0pt,minimum size=16pt] (c13) at (-1.5,1.5) {};
	\node[circle,draw=black,  inner sep=0pt,minimum size=16pt] (c14) at (-1,2) {};
	
	\node[circle,draw=black, fill=blue,   inner sep=0pt,minimum size=16pt] (c21) at (1,-1) {};
	\node[circle,draw=black, inner sep=0pt,minimum size=16pt] (c22) at (2,-1) {};
	\node[circle,draw=black, inner 	sep=0pt,minimum size=16pt] (c23) at (1.5,-1.5) {};
	\node[circle,draw=black, inner sep=0pt,minimum size=16pt] (c24) at (1,-2) {};
	
	\node[circle,draw=black, fill=blue,   inner sep=0pt,minimum size=16pt] (c31) at (-1,-1) {};
	\node[circle,draw=black, inner sep=0pt,minimum size=16pt] (c32) at (-2,-1) {};
	\node[circle,draw=black, inner sep=0pt,minimum size=16pt] (c33) at (-1.5,-1.5) {};
	\node[circle,draw=black,  inner sep=0pt,minimum size=16pt] (c34) at (-1,-2) {};
	
	\draw[line width=0.5mm, blue ]  (c11) -- (c1);\draw[line width=0.5mm, blue ]  (c1) -- (c21);\draw[line width=0.5mm, blue ]  (c11) -- (c31);
	\draw[line width=0.5mm, blue ]  (c21) -- (c31);\draw[line width=0.5mm, blue ]  (c31) -- (c1);\draw[line width=0.5mm, blue ]  (c21) -- (c11);
	\draw  (c11) -- (c12);\draw  (c11) -- (c13);\draw  (c11) -- (c14);\draw  (c13) -- (c14);\draw  (c12) -- (c13);\draw    (c12) to[out=90,in=180] (c14);
	\draw  (c21) -- (c22);\draw  (c21) -- (c23);\draw  (c21) -- (c24);\draw  (c23) -- (c24);\draw  (c22) -- (c23);\draw    (c22) to[out=270,in=0] (c24);
	\draw  (c31) -- (c32);\draw  (c31) -- (c33);\draw  (c31) -- (c34);\draw  (c33) -- (c34);\draw  (c32) -- (c33);\draw    (c32) to[out=270,in=180] (c34);
	\draw  (c1) -- (c2);\draw  (c1) -- (c3);\draw  (c1) -- (c4);\draw  (c3) -- (c4);\draw  (c2) -- (c3);\draw    (c2) to[out=90,in=0] (c4);
	
			\draw[dotted,line width=0.5mm] (-1.5,1.5) circle (1.2);
			\draw[dotted,line width=0.5mm] (1.5,1.5) circle (1.2);
			\draw[dotted,line width=0.5mm] (1.5,-1.5) circle (1.2);
			\draw[dotted,line width=0.5mm] (-1.5,-1.5) circle (1.2);
	\end{tikzpicture}
			\caption{A 3-connected extended star graph with $\BB_3=\frac{r(r-1)}{2}$ and $\CC_3=r$.}
	\label{fig:example4}
\end{center}
\end{figure}

For $k$-integrated graphs for $k \geq 3$, $\CC_k$ remains equal to $r$ whereas $\BB_k$ keeps decreasing up to $r-1$. Figure \ref{fig:example} in the Introduction shows how to construct the corresponding $k$-integrated graphs. When $k=r+1$, if there exist $r-1$ bridges connecting the $r$ central nodes, there is a path from any two central nodes with length at most $r-1$. In an extended star graph, this implies that the maximum distance between any two nodes is $r+1$, and thus the graph becomes $(r+1)$-integrated, as depicted in subfigure \ref{fig:7}.

\end{proof}

\section*{Acknowledgements}

I am indebted to an anonymous referee for helpful comments and to Sarah Fox and Vanessa Schöller for proofreading the paper.

\end{document}